\documentclass[format=sigconf, review=false, anonymous=false, dvipsnames]{acmart}

\copyrightyear{2025} 
\acmYear{2025} 
\setcopyright{acmlicensed}\acmConference[CUI '25]{Proceedings of the 7th ACM Conference on Conversational User Interfaces}{July 8--10, 2025}{Waterloo, ON, Canada}
\acmBooktitle{Proceedings of the 7th ACM Conference on Conversational User Interfaces (CUI '25), July 8--10, 2025, Waterloo, ON, Canada}
\acmDOI{10.1145/3719160.3736608}
\acmISBN{979-8-4007-1527-3/2025/07}

\usepackage{booktabs} %
\usepackage{subcaption}

\usepackage{exii-macros}

\usepackage{booktabs}

\newcommand{\rs}[1]{\textit{r\textsubscript{s}} = #1}

\hidecomments

\showrevisions{REVISIONGREEN}

\makeatletter
\g@addto@macro\normalsize{%
  \setlength\abovedisplayshortskip{-9pt}
  \setlength\belowdisplayshortskip{3pt}
}
\makeatother

\begin{document}

\tolerance=400 

\title[Psychological Ownership when Writing with AI]{Writing with AI Lowers Psychological Ownership, but Longer Prompts Can Help}

\author{Nikhita Joshi}
\orcid{0000-0001-9493-7926}
\affiliation{%
  \institution{Cheriton School of Computer Science\\University of Waterloo}
  \city{Waterloo, Ontario}
  \country{Canada}
}
\email{nvjoshi@uwaterloo.ca}

\author{Daniel Vogel}
\orcid{0000-0001-7620-0541}
\affiliation{%
  \institution{Cheriton School of Computer Science\\ University of Waterloo}
    \city{Waterloo, Ontario}
  \country{Canada}
}
\email{dvogel@uwaterloo.ca}

\renewcommand{\shortauthors}{Joshi and Vogel}

\begin{abstract}
The feeling of something belonging to someone is called ``psychological ownership.'' A common assumption is that writing with generative AI lowers psychological ownership, but the extent to which this occurs and the role of prompt length are unclear. We report on two experiments to examine the relationship between psychological ownership and prompt length. Participants wrote short stories either completely by themselves or wrote prompts of varying lengths. Results show that when participants wrote longer prompts, they had higher levels of psychological ownership. Their comments suggest they thought more about their prompts, often adding more details about the plot. However, benefits plateaued when prompt length was 75-100\% of the target story length. To encourage users to write longer prompts, we propose augmenting the prompt submission button so it must be held down a long time if the prompt is short. Results show that this technique is effective at increasing prompt length.
\end{abstract}



\begin{teaserfigure}
\centering
  \includegraphics[width=\textwidth]{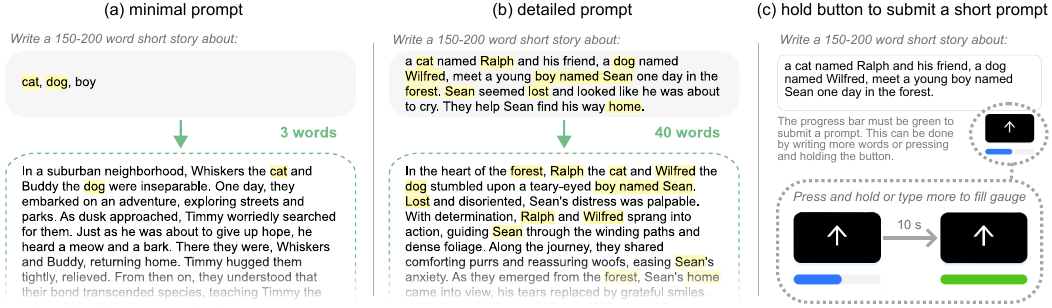}
  \caption{Writing a short story with (a) a minimal prompt with only 3 words and (b) a detailed prompt with 40 words. Yellow highlight shows similarities between the original prompt and the resulting story. Our work shows that longer prompts increase psychological ownership, likely due in part to this increased similarity. To encourage users to write longer prompts, we propose (c) a time delay experienced when pressing the button to submit a short prompt.}
  \Description{A minimal prompt, a detailed prompt, and the generated output for both. The top shows the prompts entered by the user and underneath are generated stories. Words and phrases that are similar across the prompts and generated stories are highlighted. The minimal prompt only has two similar words while the detailed prompt has over 10 words. Beside is a text box with a prompt typed in and a button. Underneath the button is a progress bar indicating how many words have been typed, which can be filled by pressing and holding the button.}
  \label{fig:teaser}
\end{teaserfigure}

\maketitle

\section{Introduction}
People often write collaboratively, which can lead to many benefits, like improved text quality \cite{Olson2017Balanced, Yim2017BalancedCollabClassroom}. However, it can also lead to negative outcomes, such as a loss of \emph{psychological ownership} \cite{Caspi2011Collaboration, Blau2009GoogleDocs, blau2009GoogleDocsSharing}. This is a concept about feelings of the text belonging to the writer, regardless of legal ownership \cite{Pierce2003PsychologicalOwnership}. Preserving feelings of psychological ownership is important when writing collaboratively, as prior work suggests that losing too much psychological ownership can deter people from writing together in the future \cite{Caspi2011Collaboration}.

People are increasingly writing ``collaboratively'' with generative AI services like ChatGPT. For example, author Rie Kudan, winner of the 170th Akutagawa Prize in Japan, used ChatGPT to write her award-winning book \cite{CTVRieKudanArticle}. Given this new form of collaborative writing, understanding how psychological ownership is affected by generative AI writing assistants is crucial. It seems intuitive that writing with a generative AI assistant will result in lower feelings of psychological ownership than writing something independently, which has been corroborated by early work investigating the roles and designs of AI writing assistants (e.g., \cite{li2024valueAIOwnership, Lehmann2022SuggestionList, Draxler2024GhostwriterOwnershipStudy}). However, we argue that more nuance of understanding is needed. Unlike collaboratively writing with another person, whose actions cannot be controlled, someone collaborating with a generative AI assistant can exert more control over the output through their prompts, which can vary in \emph{length}. Someone may write a short prompt that lacks detail about the content or style, or may write a long prompt to include these details. Writing longer  detailed prompts should result in generated text that is more representative of these additional details, and including these details requires an investment of time, energy, and sense of self; all of which are important for psychological ownership \cite{Pierce2003PsychologicalOwnership, Wang2006PsychologicalOwnershipDigital, Pierce2001PsychologicalOwnership}. Therefore, someone who invests more by writing longer prompts may have increased feelings of psychological ownership than someone who writes less, and it is possible that writing longer prompts may even lead to similar levels of psychological ownership as independent writing. Despite the potential influence of prompt length on psychological ownership, this has not been empirically investigated.

We conducted two experiments to answer the following research question: \textit{how much psychological ownership do people feel over writing produced by generative AI, and how does prompt length affect it?} During the first experiment, participants wrote 150-200-word short stories either completely by themselves or by writing prompts of varying lengths enforced through word minimums and maximums. Results showed that as participants wrote more to meet higher word minimums, they had higher levels of psychological ownership. Specifically, participants included more details about the story plot in their prompts, which resulted in more similarities between their prompts and their stories, and required more thought. The second experiment used even higher word minimums, yet psychological ownership plateaued.

These experiments required users to write longer prompts using word minimums. Though valuable for increased experimental control, this approach may not work well in practice, where users desire some flexibility. As such, we examine an important follow-up research question: \emph{how can we encourage users to write longer prompts?} We propose a simple change to the prompt entry interface, introducing a \emph{time delay} based on prompt length. Specifically, the user must press and hold the prompt submission button for some time before their prompt is submitted. The time delay is a function of prompt length, meaning if someone writes a longer prompt, they experience less delay. Using a similar creative writing task, we conducted an experiment with 20- and 60-second delays, and found prompt length increased compared to having no delays.

The effects of writing with a generative AI assistant on psychological ownership is important for the HCI community to understand, as the community is developing writing systems with generative AI support that offload work from the user. While this may be beneficial for productivity, users may come to feel negatively about such systems and stop using them over time, especially for tasks where the end product should be ``theirs.'' Our work contributes empirical and quantitative findings on the effects of writing with a generative AI assistant on psychological ownership and how this is impacted by prompt length, and a simple yet effective approach for increasing prompt length through time delays that can be integrated into current chat-based AI interfaces.

\section{Background and Related Work}
It has been long understood that people feel possessive towards objects, ideas, places, and creations. In psychology, these possessive feelings are referred to as \textit{psychological ownership} \cite{Pierce2001PsychologicalOwnership, Pierce2003PsychologicalOwnership}. Many factors contribute towards increased feelings of psychological ownership, like the investment of time, energy, and sense of self, which may explain why people feel possessive towards creative work \cite{Pierce2001PsychologicalOwnership, Pierce2003PsychologicalOwnership, Wang2006PsychologicalOwnershipDigital}. Feeling psychological ownership can have positive outcomes, like an increased sense of responsibility and citizenship behaviour \cite{Pierce2003PsychologicalOwnership}, but can negatively impact collaborative tasks \cite{Pierce2001PsychologicalOwnership}.
People often write collaboratively, which can improve text quality \cite{Olson2017Balanced, Yim2017BalancedCollabClassroom}, however, writers often feel psychological ownership over jointly-written text \cite{Caspi2011Collaboration, Blau2009GoogleDocs, Halfaker2009WikipediaOwnership, ThomSantelli2009TerritorialWikipedia}. Although some types of collaborative writing can improve psychological ownership, such as synchronous collaborations with more communication \cite{Birnholtz2013Writing}, fears of losing psychological ownership can often lead to territorial behaviours \cite{LarsenLedet2019Territorial} or avoiding future collaboration \cite{Caspi2011Collaboration}.

Psychological ownership has been identified as important to the design \cite{lee2024design} and evaluation \cite{shen2023parachute} of AI writing systems. Biermann et al. \cite{Biermann2022OwnershipLLM} conducted interviews with hobbyist and professional writers to better understand how they write and the challenges they face while writing. Participants commented on mock-up user interfaces designed for different AI assistant roles, such as using the AI to receive suggestions for new text or to generate dialog for different characters. Psychological ownership was an important factor for these writers, and some designs were poorly received because they \textit{``overstep the boundaries''} of the human writer. Draxler et al. \cite{Draxler2024GhostwriterOwnershipStudy} asked participants to write postcards entirely by themselves or to edit or choose postcards pre-written by AI before ``signing'' the postcard. Participants did not feel psychological ownership towards text that was pre-written by AI, however, they still did not list ``AI'' as an author, suggesting the presence of an \textit{AI Ghostwriter Effect}. Li et al. \cite{li2024valueAIOwnership} asked participants to write persuasive essays and stories with an AI assistant, where the human and the AI could be the primary writer or the editor. Having the AI as the primary writer lowered psychological ownership, however, participants still valued AI assistance as they were willing to forego some payment to receive help. These early evaluations suggest that psychological ownership is lowered when collaborating with AI, however, they focused on the role of the AI and did not explore other factors that contribute towards lower psychological ownership in depth.

There are many writing systems that leverage AI and LLMs (e.g., \cite{Yuan2022Wordcraft, huang2023inspo}), but we focus on those that evaluated psychological ownership. Dramatron \cite{Mirowski2023Dramaton} helps writers create theatre screenplays and scripts, but writers reported low feelings of psychological ownership over the resulting pieces. GhostWriter \cite{yeh2024ghostwriterSystem} allows writers to refine writing style preferences through highlights, but psychological ownership was the most varied and lowest scoring metric. Metaphoria \cite{Gero2019Metaphoria} suggests metaphors while writing poems, but writers felt less psychological ownership when using it, especially when \textit{``the suggestions were particularly good.''} The type of suggestion may also play a role. For example, Dhillon et al. \cite{dhillon2024ScaffoldingOwnership} found that psychological ownership decreased when writers received entire paragraphs or sentences as suggestions from an AI writing assistant. Lee et al. \cite{Lee2022CoAuthor} suggested that as writers receive more suggestions from AI, they write less and feel less psychological ownership. However, they note that these results were preliminary and more research is needed to further validate these findings.

It has been hypothesized that increased control is desirable for human-AI collaborations (e.g., \cite{Oh2018Duet, Roy2019Controllability, Rezwana2023CoCreateAI}), and that having more control can increase feelings of psychological ownership while writing. For example, Lehmann et al. \cite{Lehmann2022SuggestionList} compared AI-generated suggestions that the user had to manually accept to suggestions that were automatically inserted into the text body. They found that manually-accepted suggestions led to higher feelings of psychological ownership, which may be due to the increased control at the composition stage.

\medbreak

To summarize, prior work has shown that psychological ownership is important to writers who write collaboratively with humans and with AI. Although some AI writing systems have included features to give writers more control to increase psychological ownership, the effects of more fundamental properties, such as the AI prompt length, have yet to be explored.

\section{Experiment 1: Prompt Length}
The goal of this experiment is to understand the impact of writing with generative AI on psychological ownership and how prompt length affects it. Through a within-subjects experimental design, participants wrote short stories with generative AI assistance by writing prompts of varying word lengths, which were enforced through word minimums and maximums for increased experimental control: 3 words, 50-100 words, and 150-200 words. As a baseline, participants also wrote a 150-200 word short story without any AI assistance.

\subsection{Participants}
We recruited 41 participants through the crowdsourcing experiment service, Prolific.\footnote{\href{https://www.prolific.com}{https://www.prolific.com}} Participants were restricted to the United States and Canada, those who had completed at least 2,500 tasks, and those who had an approval rating greater than 98\%. We manually examined all open-ended responses and user input to identify fraudulent responses \cite{Ryan2020MTurkFraud} and filtered out participants who experienced technical issues with our interface and those who cheated by padding out their prompts or stories with dummy text to meet the word minimums and those who did not make enough keystrokes to generate their user input, suggesting they copied and pasted content from another source.
In total, 10 participants (24\%) were excluded, leaving 31 valid responses (Table \ref{tab:exp1Demo}). All self-reported being proficient at reading and writing in English (all $\geq$ 5 on a 1-7 scale). Participants received \$15 as remuneration.

\subsection{Apparatus and Task}
\begin{figure*}[bt]
	\centering
	\includegraphics{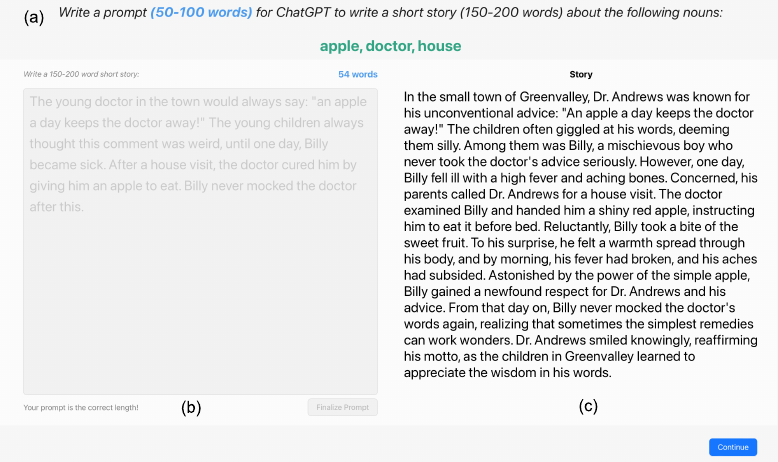}
	\caption{Experimental interface (shown with AI assistance): (a) top toolbar showing the instructions and three nouns the story had to be about; (b) a text box where the participant wrote their prompt or short story; (c) the story written by themselves or with AI assistance.}
    \Description{Top toolbar showing instructions ("write a prompt (50-100 words) for ChatGPT to write a short story (150-200 words) about the following nouns: apple, doctor, house"). A text box on the left shows a 56 word story entered by the user. The right shows the story generated by AI. A Continue button on the bottom of the page is enabled.}
	\label{fig:exp1-task} %
\end{figure*}

The experiment software was a Node.js and React web application (Figure \ref{fig:exp1-task}). The main writing interface had a toolbar on the top and a side-by-side view underneath. The top toolbar contained instructions for the trial and three nouns that the story had to be about (i.e., a unique and randomly assigned ``triad'' from a set of nine triads adapted from Foley et al.'s text composition task \cite{Foley2020Composition}), which is useful for encouraging creative thinking on the spot during experiments \cite{dunlop2016Images}. Writing independent short stories is an ideal task for this type of experiment as creative writing requires a high personal investment and expression of self, which are important for psychological ownership, yet it provides increased control for an experiment.

A text box on the left provided a space for participants to write a prompt for the GPT-3.5 Turbo model. In a baseline condition, they wrote a complete story. Copying and pasting within the text box were disabled to prevent cheating, but this may not work across all browsers, so keystroke-level activity was also logged to verify response validity. A word counter above the text box displayed the number of words in the text box. Once the word count satisfied the requirements for the condition (i.e., the word count was between the minimum and maximum), a \textit{Finalize} button could be pressed, causing the story to ``type out'' on the right side of the screen, following the behaviour of LLM AI systems like ChatGPT. The \textit{Finalize} button could only be pressed once. The story was either generated by the model by including the participant's prompt within an engineered prompt,\footnote{The prompt used to generate the story is included in the supplementary materials.} or it was the same text written by the participant in the baseline. The same typing effect was used in both cases, and this provided an opportunity for participants to read the story. After the story was fully typed, a \textit{Continue} button could be pressed to end the trial.

\subsection{Procedure}

Participants received a link to the web application through Prolific and the experiment was restricted to laptop and desktop devices. After entering basic demographic information and reading instructions, they were shown the writing interface and were asked to write a short story or a prompt about the three presented nouns. After the writing stage, they answered questions about their story and the writing experience. They repeated this for all conditions before answering questions about their overall thoughts and preferences. The entire experiment took approximately one hour.

\subsection{Choice of Prompt Lengths}
We were interested in examining the effects of prompt length, believing that being required to write longer prompts will increase the prompt level of detail and improve psychological ownership. As such, we needed to test a range of prompt lengths representing key cases, like a story generated with very little guidance from the writer, a story heavily guided by the writer, and a story written solely by the writer. At the same time, we had to ensure that the experiment duration was not too long for participants. Therefore, our choice of word minimums (and word maximums) was guided by the duration of conditions requiring more writing from the participant (e.g., writing a story without any AI assistance). We ultimately decided that the final stories for all conditions had to be 150-200 words, so that the experiment was not too time-consuming but the stories were still long enough to be fully developed.
With a 150-200 word final story in mind, we chose conditions to examine the effects of prompt length and a baseline:

\begin{itemize}
    \item \textbf{3 words with GPT}, representing 1\% of the maximum final story word count. This best approximates the case where the story is fully generated by an AI assistant.
    \item \textbf{50-100 words with GPT}, representing 25-50\% of the maximum final story word count. The story is generated by an AI assistant but has some guidance from the participant.
    \item \textbf{150-200 words with GPT}, representing 75-100\% of the maximum final story word count. The story is generated by an AI assistant, but with a lot of guidance from the participant.
    \item \textbf{150-200 words without GPT}, a baseline representing the case where the story is fully written by the participant.
\end{itemize}

\subsection{Design}
This is a within-subjects design with one primary independent variable, \f{condition} (levels: \f{ai-3}, \f{ai-50}, \f{ai-150}, \f{no-ai}; the number representing the word minimum). \f{condition} was randomly assigned.
For each \f{condition}, the primary measures were 10 subjective question scores, all interval numeric scores within a 1-7 range.\footnote{See the supplementary materials for question wording and data from the experiment.} Four questions represented metrics related to psychological ownership and the phrasing was similar to prior work \cite{nicholes2017measuring, Caspi2011Collaboration}: \m{Personal Ownership}, \m{Responsibility}, \m{Personal Connection}, \m{Emotional Connection}. We average these four scores to create a composite measure, \m{Psychological Ownership}, which is a common technique in psychology papers. The internal consistency reliability was calculated using Cronbach's alpha, and the reliability was very high ($\alpha$ = 0.96), indicating that \m{Psychological Ownership} was a reliable composite measure for data analysis.
Six questions represented metrics from the NASA-TLX: \m{Mental Demand}, \m{Physical Demand}, \m{Temporal Demand}, \m{Performance}, \m{Effort}, and \m{Frustration}.
In addition, objective metrics were calculated from logs: the time taken in minutes to write the prompt or story (\m{Duration}); the number of words in each (\m{Word Count}); and the semantic text similarity between the final story and the participant's prompt (\m{Text Similarity}), which was calculated using Google's Universal Sentence Encoder (0-1 range; 1 being identical texts) \cite{cer2018universalSentenceEncoder}.

\NewDocumentCommand\friedmanN{ m m m m g }{%
    \IfNoValueTF{#5}{%
         {\small$\chi^2_{#1, N=#2}=#3$, $p<#4$}%
    }{%
         {\small$\chi^2_{#1, N=#2}=#3$, $p<#4$, $W=#5$}%
    }%
}

\NewDocumentCommand\friedmanEta{ m m m m g }{%
    \IfNoValueTF{#5}{%
         {\small$\chi^2_{#1, N=#2}=#3$, $p<#4$}%
    }{%
         {\small$\chi^2_{#1, N=#2}=#3$, $p<#4$, $\eta^2=#5$}%
    }%
}

\NewDocumentCommand\ppquote{ m g }{%
    \IfNoValueTF{#2}{%
         \textit{``#1''}%
    }{%
         \textit{``#1''}{\,}{\small(#2)}%
    }%
}

\newcommand{\iqr}[1]{$\textsc{iqr}\!=\!#1$}

\section{Experiment 1: Results}
Where applicable, we use a Friedman test and Wilcoxon signed-rank tests, with Holm's corrections for multiple comparisons; and Spearman's correlations. \textit{To streamline the presentation of results, all statistical test details are shown in Appendix \ref{section:appendixStudy1} in Table \ref{tab:exp1Significance}}.
Where applicable, graphs show the mean and individual participant scores, and all error bars represent 95\% confidence intervals (bootstrapped with 10,000 resamples).

\subsection{Psychological Ownership}
\begin{figure}[h]
	\centering
	\includegraphics[width=0.47\textwidth]{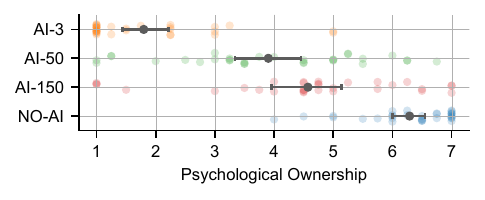}
	\caption{\m{Psychological Ownership} by \f{condition}. Higher scores correspond to higher feelings of \m{Psychological Ownership}.}
    \Description{Confidence intervals and individual participant data for Psychological Ownership. As the prompt length increases, Psychological Ownership increases. The baseline, NO-AI has its entire confidence interval greater than 6.}
	\label{fig:psychologicalOwnership} %
\end{figure}

Overall, writing more words improved feelings of psychological ownership (Figure \ref{fig:psychologicalOwnership}). A significant effect of \f{condition} on \m{Psychological Ownership} and post hoc tests revealed that \f{ai-3} led to the lowest scores (\mean{1.80}, \sd{1.13}), followed by \f{ai-50} (\mean{3.90}, \sd{1.61}), \f{ai-150} (\mean{4.57}, \sd{1.70}), and \f{no-ai} (\mean{6.29}, \sd{0.80}). 

\subsection{Writing Activities}
To better understand factors that may have contributed to these results, we examine participants' writing activities and task workload. Open-ended responses were grouped by the first author as the data was straightforward given the number of participants and the short response length (2-3 sentences each).

\subsubsection{Duration}

\begin{figure}[h]
	\centering
	\includegraphics[width=0.47\textwidth]{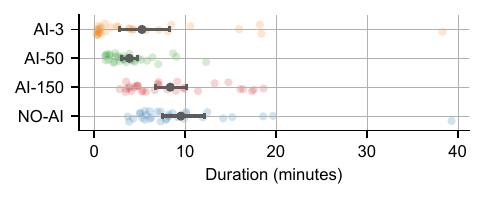}
	\caption{\m{Duration} by \f{condition}.}
    \Description{Confidence intervals and individual participant data for Duration. Most participants took less than 20 minutes.}
	\label{fig:duration} %
\end{figure}

We anticipated that spending more time writing may lead to higher \m{Psychological Ownership}, but they were only moderately positively correlated (\rs{.45}; \p{.001}). A significant effect of \f{condition} on \m{Duration} and post hoc tests revealed that \f{ai-3} (\mean{5.26}, \sd{8.08}) and \f{ai-50} (\mean{3.88}, \sd{2.60}) were faster than \f{ai-150} (\mean{8.37}, \sd{4.91}) and \f{no-ai} (\mean{9.54}, \sd{6.83}; Figure \ref{fig:duration}).

\subsubsection{Word Count}
\begin{figure}[bt]
	\centering
	\includegraphics[width=0.47\textwidth]{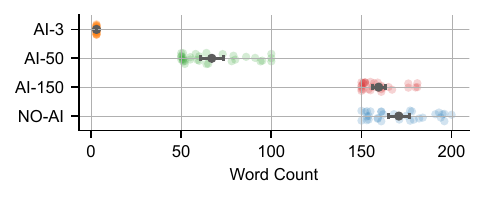}
	\caption{\m{Word Count} by \f{condition}.}
    \Description{Confidence intervals and individual participant data for Word Count. As expected, participants wrote more words as prompt length increased. Comparing AI-150 and NO-AI shows that most data points are clustered around the 150 word minimum for AI-150, but the points are more spread out for NO-AI.}
	\label{fig:wordCount} %
\end{figure}

We observed a strong positive correlation between \m{Word Count} and \m{Psychological Ownership} (\rs{.72}; \p{.001}), suggesting that as participants wrote more, they felt more psychological ownership over their writing.
As expected, we observed a significant effect of \f{condition} on \m{Word Count} with post hoc tests showing participants wrote fewer words with \f{ai-3} and \f{ai-50} (\mean{66.94}, \sd{17.91}) than they did with \f{ai-150} (\mean{159.52}, \sd{9.86}) and \f{no-ai} (\mean{170.74}, \sd{16.63}; Figure \ref{fig:wordCount}).
Despite having the same word limits, we also observed a significant difference between \f{ai-150} and \f{no-ai}. One possibility is that participants encountered a mental block with \f{ai-150}, as some mentioned how it was \ppquote{more taxing to write [a] longer prompt}{P31} 
and how they \ppquote{couldn't think of what [...] to write about}{P21}.

\subsubsection{Text Similarity}
\begin{figure}[h]
	\centering
	\includegraphics[width=0.47\textwidth]{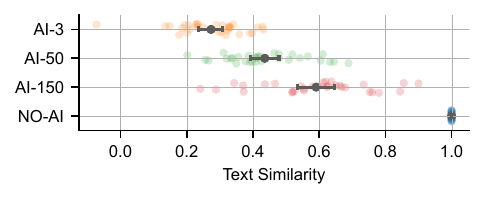}
	\caption{\m{Text Similarity} by \f{condition}. Higher scores refer to more similarities between the input and output (1 = identical texts).}
    \Description{Confidence intervals and individual participant data for Text Similarity. As the prompt length increases, Text Similarity increases.}
	\label{fig:textSimilarity} %
\end{figure}

Overall, writing longer prompts led to more semantic text similarity with the generated story. There was a strong positive correlation between \m{Word Count} and \m{Text Similarity} (\rs{.82}, \p{.001}), and a significant effect of \f{condition} on \m{Text Similarity}, revealing that prompts written with \f{ai-3} were the least similar to the final story (\mean{0.27}, \sd{0.10}), followed by \f{ai-50} (\mean{0.43}, \sd{0.12}), \f{ai-150} (\mean{0.59}, \sd{0.16}), and \f{no-ai} (Figure \ref{fig:textSimilarity}). Much like \m{Word Count}, having increased \m{Text Similarity} was strongly positively correlated with higher \m{Psychological Ownership} (\rs{.75}, \p{.001}).

\subsubsection{Prompt Strategies}
We observed participants adopting several strategies with their prompts (Table \ref{tab:prompts}). As expected, most participants wrote the three specified nouns for \f{ai-3}, so we only consider \f{ai-50} and \f{ai-150}. Only 4 prompts (6\%) purely consisted of instructions without any details about the story plot; 25 (40\%) purely consisted of story plot details and resembled partially complete stories; and majority (33, 53\%) were a mix of instructions and story details.
As would be expected, it seems like requiring participants to write longer prompts encouraged greater levels of details about the story, something that 11 (35\%) participants noted, for example: \ppquote{when I provide more words, I see that ChatGPT produces a story that is almost the same as mine}{P1}. Of these participants, 8 (25\%) noted how this increased feelings of psychological ownership, for example: \ppquote{with longer prompts, I could influence what was on display and felt more ownership [towards] the work}{P13}.

\begin{table*}[t]
    \centering
    \caption{Example prompt types submitted by participants (examples from \f{ai-50} condition).}
    \small %
\begin{tabular}{p{0.3\textwidth}|p{0.3\textwidth}|p{0.3\textwidth}}
\toprule
\textbf{Plot} & \textbf{Mix} & \textbf{Instructions} \\
\midrule
\ppquote{The girl asked her grandma for a dog every time she saw her. The grandma kept saying no, until finally she surprised her granddaughter with a dog for Christmas that year. The girl had tears of joy that she finally got a dog and was in disbelief that it was happening.}{P18} & \ppquote{Create an average word paragraph using the following nouns: House, grandpa and mug. Start with a setting of grandpa sitting in a rocking chair on the porch. Have grandpa tell a quick story to his 8 year old grand daughter named Lisa. End story with grandpa going into the house.}{P15} & \ppquote{Write a short story about these three nouns: Boy, boat, cat. The story should be 150 - 200 words in length. The whole story should be about these three nouns. Write a story that can be read to children, so the words should be as simple as possible so children could understand.}{P26} \\
\bottomrule
\end{tabular}
    \Description{Example prompt types: plot, mix, and instructions. The plot prompt does not contain any instructions while the mix contains some plot elements and some instructions. The instructions prompt does not contain any plot details beyond the three nouns they were asked to write about.}
    \label{tab:prompts}
\end{table*}

\subsubsection{Task Workload}
\begin{figure}[h]
	\centering
	\includegraphics[width=0.47\textwidth]{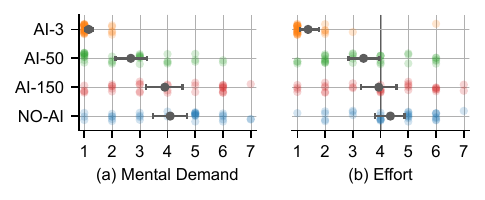}
	\caption{(a) \m{Mental Demand} and (b) \m{Effort} by \f{condition}. Lower scores correspond to lower \m{Mental Demand} and \m{Effort}.}
    \Description{Confidence intervals and individual participant data for Mental Demand and Effort. As the prompt length increases, Mental Demand and Effort generally increase, however, there is not much difference between AI-150 and NO-AI.}
	\label{fig:mentalEffort} %
\end{figure}

The longest prompt and writing the full story were more mentally demanding and the shortest prompt required the least effort (Figure \ref {fig:mentalEffort}), but we did not observe meaningful trends for other task workload factors.
For \m{Mental Demand}, a significant effect of \f{condition} and post hoc tests revealed that \f{ai-3} (\mean{1.61}, \sd{0.37}) and \f{ai-50} (\mean{2.68}, \sd{1.62}) were less mentally demanding than \f{ai-150} (\mean{3.90}, \sd{1.90}) and \f{no-ai} (\mean{4.10}, \sd{1.81}; Figure \ref{fig:mentalEffort}a). There were no significant differences between \f{ai-150} and \f{no-ai}. \m{Mental Demand} was strongly positively correlated with \m{Psychological Ownership} (\rs{.65}, \p{.001}). 

There was a significant effect of \f{condition} on \m{Effort}. Unsurprisingly, post hoc tests showed that \f{ai-3} required the least \m{Effort} (\mean{1.39}, \sd{0.99}). Although \f{ai-50} (\mean{3.39}, \sd{1.63}) required less effort than \f{no-ai} (\mean{4.35}, \sd{1.52}), there were no significant differences between \f{ai-50} and \f{ai-150} (\mean{3.94}, \sd{1.88}) or between \f{ai-150} and \f{no-ai} (Figure \ref{fig:mentalEffort}b). \m{Effort} was strongly positively correlated with \m{Psychological Ownership} (\rs{.66}, \p{.001}).
The relationship between \m{Mental Demand} and \m{Effort} and \m{Psychological Ownership} makes sense; it has been long understood that investing more energy and self improves psychological ownership \cite{Pierce2001PsychologicalOwnership, Pierce2003PsychologicalOwnership, Wang2006PsychologicalOwnershipDigital}, which requires more thought and effort. 

\subsection{Summary}
To summarize, our results suggest that requiring people to write longer prompts increases psychological ownership for their writing. By writing more, prompts became more similar to the resulting stories, likely because writers were encouraged to include more details about the story and its plot, which required more mental demand. An interesting question is: does this effect plateau with even longer prompts, perhaps even longer than the generated text?

\section{Experiment 2: Plateau}
To better understand this, we conducted a follow-up experiment with 39 new participants. Five (13\%) were removed from the analysis for cheating or because they experienced technical issues, leaving 34 participants (Table \ref{tab:exp2Demo}). All participants self-reported being proficient at reading and writing in English (all $\geq$ 5 on a 1-7 scale). The protocol and measures were the same as Experiment 1 ($\alpha$ = 0.92 for \m{Psychological Ownership}), but participants had to write prompts that were 150-200, 175-200, and 200-250 words long. The experiment was slightly longer at roughly 90 minutes and participants received \$22.50 as remuneration.

\subsection{Choice of Prompt Lengths}
Two conditions were identical (\f{ai-150} and \f{no-ai}), and two new word minimums and maximums pushed participants to write even more:

\begin{itemize}
    \item \textbf{175-200 words with GPT}, representing 87.5-100\% of the maximum final story word count. 175 words roughly corresponds to the mean word count of solo-writing in Experiment 1 (\f{ai-175}), which encourages participants to write slightly more words than what was anticipated for \f{no-ai}.
    \item \textbf{200-250 words with GPT}, representing 100-150\% of the maximum final story word count. This condition requires the participant to write more words than the baseline solo-writing condition and the generated story, which may lead to higher feelings of psychological ownership (\f{ai-200}).
\end{itemize}

\subsection{Results}
We use the same analysis as Experiment 1 with \textit{detailed statistical test results in Table \ref{tab:exp2Significance}
in Appendix \ref{section:appendixStudy2}}.

\subsubsection{Psychological Ownership}
\begin{figure}[t]
	\centering
	\includegraphics[width=0.47\textwidth]{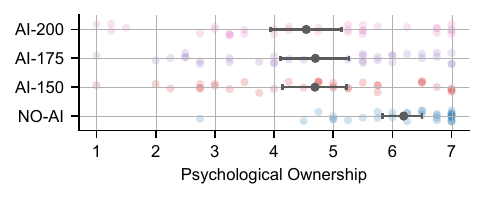}
	\caption{\m{Psychological Ownership} by \f{condition}. Higher scores correspond to higher feelings of \m{Psychological Ownership}.}
    \Description{Confidence intervals and individual participant data for Psychological Ownership. The confidence intervals and spread of data are similar for AI-200, AI-175, and AI-150, but NO-AI has a tighter spread of data and higher confidence intervals.}
	\label{fig:psychologicalOwnershipExp2} %
\end{figure}

Overall, psychological ownership when writing with AI seems to plateau beyond a 150 word minimum (Figure \ref{fig:psychologicalOwnershipExp2}). There was a significant effect of \f{condition} on \m{Psychological Ownership}, and post hoc tests revealed that people felt less \m{Psychological Ownership} for all conditions that involved AI than \f{no-ai} (\mean{6.19}, \sd{0.99}), and there were no significant differences between \f{ai-150} (\mean{4.69}, \sd{1.64}), \f{ai-175} (\mean{4.70}, \sd{1.73}), and \f{ai-200} (\mean{4.54}, \sd{1.81}).

\subsubsection{Writing Activities}
This plateauing effect is further supported by participants' writing activities. Although \m{Word Count}, \m{Text Similarity}, \m{Mental Demand}, and \m{Effort} were all strongly positively correlated with \m{Psychological Ownership} in Experiment 1, none of these correlations held in this experiment (0.22 $\leq$ \textit{r\textsubscript{s}} $\leq$ 0.48 for significant correlations). Furthermore, there were no meaningful significant effects of \f{condition} on \m{Text Similarity}, \m{Mental Demand}, and \m{Effort}, despite significant differences in \m{Word Count}.

\subsection{Summary}

Overall, our results suggest a plateau point around a 150 word minimum for the 150-200 word stories generated in our experiment. Even when participants wrote even longer prompts with AI, they still felt less psychological ownership than writing alone. For this writing task, a prompt length similar to the target word length appears to be a ``sweet spot'' \cite{onarheim2015balancing} for trying to maximize psychological ownership when writing with AI, as writing with higher word minimums did not improve psychological ownership.

\section{Encouraging Longer Prompts in a Real Interface}
In our experiments, we had to enforce strict prompt length minimums for the purpose of internal validity and experimental control. But this kind of ``hard'' constraint may not work well in real-world applications, where users expect more flexibility. How can we encourage users to write more within a typical chat-style AI user interface?

To better understand how users can be nudged towards writing longer prompts without strict word minimums, we designed alternative interface designs and interaction techniques that modify the presentation of the output and the prompt entry interface while remaining highly compatible with current chat-based generative AI tools (Figure \ref{fig:interactions}; see Appendix \ref{section:appendixInteractionTechs} for detailed descriptions). We ran a pilot experiment (n=90, 11-18 per condition) to better understand which modifications were the most promising for increasing prompt length, and our preliminary results\footnote{Data from these participants is also included in the main results of Experiment 3.} suggested that integrating a \emph{time delay} into the prompt submission button could be effective (Figure \ref{fig:pilotExp3}). 

With the time delay modification, the user must press and hold the button for some time to fill up a ``gauge,'' only after which the prompt is submitted. However, this gauge can also be filled by writing more words in the prompt entry text box. These two methods of filling the gauge provide the user with a choice: they can either write a very short prompt and press and hold the submission button for several seconds, or they can write a longer prompt and press and hold the button for a shorter amount of time, or for no time at all if their prompt length met or exceeded a system-defined ``word minimum'' that leads to higher levels of psychological ownership.

The potential of time delays for increasing prompt length makes sense, as prior work shows that purposely slowing user interactions can effectively nudge users. Specifically, they can cause people to shift their attention and focus \cite{nielsenTimeDelays1993} and think more analytically \cite{WASON1974141, Wang2014NudgeFacebookDelay, CFFsBuccina2021}, which can lead to positive outcomes. For example, time delays can encourage users to reflect on social media posts before they are shared with the public \cite{Wang2014NudgeFacebookDelay}; help users evaluate the validity of output generated by an AI assistant \cite{CFFsBuccina2021}; and even help users learn expert menu selection techniques associated with marking menus \cite{kurtenbach1994user, Lewis2020Delays}.

Given these findings, we investigate the use of time delays in more depth to see whether they could increase prompt length. Based on prior work, we suspect that the act of pressing and holding the prompt submission button for several seconds will deter users from writing shorter prompts as they may lose attention and focus, and encourage users to reflect on their prompts and expand them.

\section{Experiment 3: Delayed Submission}
The goal of this experiment is to understand the effectiveness of a time delay experienced by pressing and holding the prompt submission button. Participants completed a similar creative writing task as the previous two experiments (writing prompts to generate a 150-200 word short story), but were assigned a random time delay condition for the entire experiment. Two baselines included: no modifications, and a strict word minimum like the previous experiments.

\subsection{Participants}
We recruited 198 new participants with 42 (21\%) removed from the analysis for cheating or experiencing technical issues, leaving 156 valid responses (Table \ref{tab:exp3Demo}). All reported being proficient at reading and writing in English (all $\geq$ 4 on a 1-7 scale). Participants received \$15 as remuneration.

\subsection{Apparatus}
\begin{figure*}[tb]
	\centering
	\includegraphics{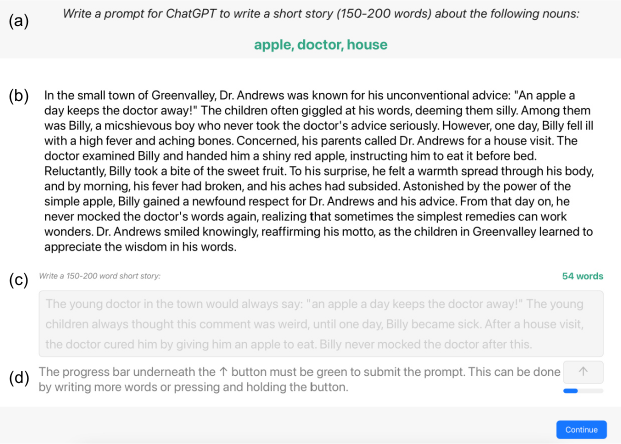}
	\caption{Experimental interface with a time delay: (a) top toolbar showing the instructions and three nouns the story had to be about; (b) the story written with AI assistance; (c) a text box where the participant wrote their prompt; (d) instructions about the technique.}
    \Description{Top toolbar showing instructions ("write a prompt (50-100 words) for ChatGPT to write a short story (150-200 words) about the following nouns: apple, doctor, house"). Below is a generated story. Below this is a text box with a prompt that is 54 words long. The submit button has a progress bar underneath. Instructions under the text box guide users ("the progress bar underneath the submit button must be green to submit the prompt. This can be done by writing more words or pressing and holding the button.")}
	\label{fig:nudgeExpInterface} %
\end{figure*}

The experiment software was similar to that used in the prior experiments, but with a few cosmetic modifications to make the interface more closely resemble existing chat-based AI user interfaces like ChatGPT (Figure \ref{fig:nudgeExpInterface}). The prompt entry and output interfaces were stacked on top of each other. The prompt entry text box, which has a default height of 46 pixels, grows by 46 pixels with every new line typed, until a maximum height of 300 pixels.

With the exception of the baseline condition with no modifications, a brief message underneath the prompt entry text box provides instructions for prompt submission (Figure \ref{fig:nudgeExpInterface}d). To mitigate risks of bias, the message simply explained how the delayed button worked. Inline with the previous experiments, the time delay mechanics used a 150-word minimum for optimal psychological ownership. The baseline condition had to indicate 150 words as the minimum, but the delayed button conditions made no mention of 150 words.

\subsubsection{Delayed Button Interaction and Mechanics}
A ``gauge'' that must be filled is visualized as a progress bar below the prompt submission button (Figure \ref{fig:teaser}c). When the button is not pressed, the length of the bar is the ratio of the current prompt word count to the \emph{word minimum} (150 words in this experiment). When the button is held down, the progress bar gradually increases (updated every 500 ms) according to an \emph{experienced delay}. This time delay period is proportional to the unfilled bar ratio to the \emph{maximum delay}. For example, consider a 150-word minimum and a maximum delay of 20 s. If the user writes a 15-word prompt, the experienced delay (the time they must hold the button down) will be 18 s. If the user enters 135 words, the experienced delay will only be 2 s. If the user enters at least 150 words, the experienced delay will be 0 s (i.e., the button works with a normal click).
The progress bar is blue when not full, turning green when enough words have been typed or when enough delay has been experienced by pressing and holding the button. Note that the user must press and hold the button in one single act; releasing it before the progress bar turns green resets the progress bar to its original size based on the number of words in the prompt.

\subsection{Procedure}
The procedure was nearly identical to the prior experiments. However, rather than experiencing all conditions once, participants experienced a single condition four times as we suspected a learning curve for the experimental conditions. As the focus of this study was not psychological ownership, they did not answer questions after every story, instead, they answered questions about their experience after all four trials. The entire experiment took 20 to 60 minutes, depending on the condition.

\subsection{Choice of Time Delays}
Prior work suggests that delays longer than 10 seconds will cause users to shift their attention \cite{nielsenTimeDelays1993}, which is desirable as participants may opt to write a longer prompt instead of pressing and holding the prompt submission button. Delays of 20 seconds are tolerable \cite{selvidge1999long}, so the initial pilot experiment described above tested a 20-second delay. The results suggested a bimodal distribution of prompt word length, participants either wrote very short prompts, or very long prompts (Figure \ref{fig:pilotExp3}).

Based on the pilot, we introduce two additional time delay conditions: an even longer, 60-second delay \cite{selvidge1999long}, to increase frustration and encourage more participants to write longer prompts; and a 0-second delay, where the same progress bar visualization was shown underneath the prompt submission button, but the participant did not have to press and hold the button. The latter was included to ensure that the effects were indeed caused by a time delay, rather than confounding factors such as minor changes to the user interface. Therefore, we test the following five conditions:

\begin{itemize}
    \item \textbf{No Modifications} (n=32), a baseline where participants did not see or experience any modifications to the user interface. This best approximates current chat-based AI interfaces.
    \item \textbf{0-Second Delay} (n=32), where the participant sees a progress bar indicating how many words out of 150 they had typed, but do not experience any delay.
    \item \textbf{20-Second Delay} (n=31), where the participant sees the same progress bar indicator and can experience a delay up to 20 seconds long, depending on how many words they write out of 150.
    \item \textbf{60-Second Delay} (n=32), which is similar to the 20-second delay condition, but with a longer, 60-second delay.
    \item \textbf{150-Word Minimum} (n=29), a baseline representing the case where the participant must write a prompt that is at least 150 words long. This condition is guaranteed to increase prompt length, which allows us to make relative comparisons to other conditions.
\end{itemize}

\subsection{Design}
This is a mixed-design with two primary independent variables: \f{condition} (levels: \f{none}, \f{0-sec}, \f{20-sec}, \f{60-sec}, \f{words}), which was between-subjects and randomly assigned, and \f{trial} (levels: \f{1}, \f{2}, \f{3}, \f{4}), which was within-subjects.

The primary measures were objective metrics calculated from logs: the word count of the submitted prompt (\m{Final Word Count}), the number of times the participant tried to submit a prompt by pressing the prompt submission button (\m{Submission Attempts}), the word count of the prompt for each submission attempt (\m{Attempted Word Count}), and the number of seconds they held the button for each submission attempt (\m{Attempted Duration}). Note that \m{Submission Attempts}, \m{Attempted Word Count} and \m{Attempted Duration} are only applicable for the \f{20-sec} and \f{60-sec} conditions, as the prompt submission button could only be pressed once for all other conditions due to the lack of delay (i.e., \m{Attempted Word Count} is equal to \m{Final Word Count} for \f{none}, \f{0-sec}, and \f{words}).

\section{Experiment 3: Results}
Where applicable, we use the Aligned Rank Transform (ART) \cite{wobbrock2011aligned} and post hoc contrast tests (ART-C) \cite{elkin2021aligned}, Kruskal-Wallis tests and Mann-Whitney U tests, Holm's corrections for multiple comparisons, and Spearman's correlations. \emph{Detailed statistical test results are in Table \ref{tab:exp3Significance}.}

\subsection{Final Word Count}
\begin{figure}[tbh]
	\centering
	\includegraphics[width=0.47\textwidth]{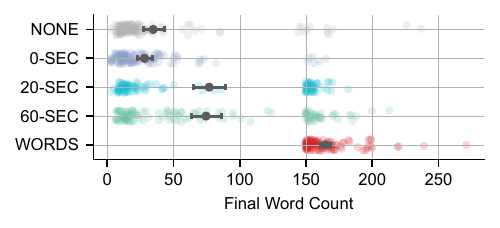}
	\caption{\m{Final Word Count} by \f{condition}.}
    \Description{Confidence intervals and individual participant data for Final Word Count. NONE and 0-SEC are similar and 20-SEC and 60-SEC are similar. 20-SEC and 60-SEC have two clusters: one between 0 and 50 words, and another around 150 words. WORDS are all above 150 words.}
	\label{fig:wordCountExp3} %
\end{figure}

Overall, experiencing a time delay when submitting a prompt increased prompt length (Figure \ref{fig:wordCountExp3}). A significant main effect of \f{condition} on \m{Final Word Count} and post hoc tests revealed that \f{20-sec} (\mean{76.91}, \sd{68.13}) and \f{60-sec} (\mean{74.57}, \sd{63.80}) led to significantly higher \m{Final Word Count} than \f{none} (\mean{34.66}, \sd{46.56}) and \f{0-sec} (\mean{28.09}, \sd{32.69}). This was a result of the time delay, rather than changes to the user interface, as there were no significant differences between \f{none} and \f{0-sec}. However, as expected, a time delay is not as effective as a strict word minimum, as \f{words} had the highest \m{Final Word Count} (\mean{164.65}, \sd{19.54}).

\subsection{Learning Effect}
We expected that participants would learn how to use the delayed prompt submission across multiple trials, but this did not occur. Rather, most participants discovered their preferred way of interacting with the button in the first trial, and continued this for all four trials. Specifically, significant main effects of \f{trial} on \m{Submission Attempts} and \m{Attempted Word Count} and post hoc tests only revealed significant differences between \f{trial} \f{1} and all other trials. Therefore, we focus our discussion of the learning effect on the results from the first trial.

\subsubsection{Submission Attempts}
Participants made significantly more \m{Submission Attempts} in the first trial for \f{20-sec} (\mean{4.35}, \sd{4.45}) and \f{60-sec} (\mean{4.28}, \sd{4.70}) than subsequent trials (1.06 $\leq \textsc{m} \leq$ 1.84; 0.24 $\leq \textsc{sd} \leq$ 4.78). However, the median \m{Attempted Duration} was only 0.1 seconds for both \f{20-sec} and \f{60-sec}, suggesting that participants were initially trying to press the button normally, without holding it down.

\subsubsection{Attempted Word Count versus Final Word Count}
Examining prompt submission attempts in the first trial, \m{Attempted Word Count} for \f{20-sec} (\mean{32.86}, \sd{42.55}) and \f{60-sec} (\mean{36.91}, \sd{43.96}) was on par with the \m{Final Word Count} of \f{none} (\mean{31.59}, \sd{42.97}) and \f{0-sec} (\mean{30.34}, \sd{40.81}), with no significant differences observed. But \f{20-sec} and \f{60-sec} led to a significantly higher \m{Final Word Count}, suggesting that participants eventually wrote longer prompts after attempting to write shorter prompts. 
Specifically, in the first trial for \f{20-sec} and \f{60-sec}, participants first tried to submit prompts that were 26.81 and 39.16 words long on average, before finally submitting prompts that were 78.10 and 69.03 words long on average, respectively. In other words, \m{Final Word Count} nearly tripled and doubled in length from the first \m{Attempted Word Count}.

\subsection{Strategies}
\begin{figure*}[tbh]
	\centering
	\includegraphics{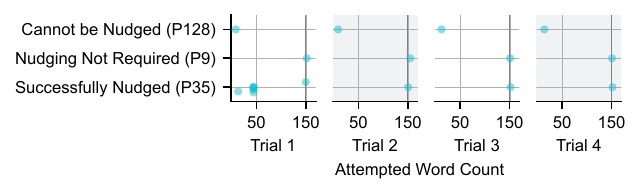}
	\caption{Prompt word count each time the submit button was pressed (i.e., \m{Attempted Word Count}) for these participants. Examples from \f{20-sec} condition, each \m{Submission Attempt} is a blue dot. }
    \Description{Three participants' Attempted Words Count across all four trials. First is a participant who cannot be nudged, who always writes short prompts (less than 20 words). Next is a participant who does not need to be nudged because their prompts are always at least 150 words long. Last is a participant who was successfully nudged. In trial 1, there were 3 Submission Attempts with Attempted Word Counts less than 50, but then the Final Word Count was 150 words. Trials 2, 3, and 4 had 1 Submission Attempt, with Final Word Count at least 150 words long.}
	\label{fig:strategiesExp3} %
\end{figure*}

Overall, there were no significant differences between \f{20-sec} and \f{60-sec} for \m{Final Word Count}, and both techniques invoked similar behaviours from participants (Figure \ref{fig:wordCountExp3}): they either wrote short prompts ($\leq$ 50 words), or long prompts ($\geq$ 150 words).
Examining individual prompt submission attempts for the first trial for \f{20-sec} and \f{60-sec} provides insights into these diverging strategies. Some participants are consistent and do not change their behaviour, with 29 participants (46\%) showing no differences between their first \m{Attempted Word Count} and \m{Final Word Count}, but 32 participants (51\%) wrote more, with 17 of them (27\%) writing at least 50 more words, which consisted of more details about the story plot and writing style.
These differences suggest three primary types of users and strategies (Figure \ref{fig:strategiesExp3}): people who do not need to be nudged as they always write long prompts; people who are more resistant to being nudged and always write shorter prompts; and people who can be nudged to write longer prompts.

Feedback from participants suggested that, as expected, those who do not need to be nudged because they wrote longer prompts did not notice any delay, with comments like \ppquote{it's nothing that caught my attention}{P9}. Participants who prefer to write shorter prompts noted that pressing and holding the button was a worthwhile trade-off, for example, \ppquote{it required far less time and effort to hold down the button than to write the stories myself}{P46}. This may be related to typing speed. A participant who wrote short prompts said \ppquote{[someone] who types fast may not mind [writing more]}{P136} while some participants said the opposite, for example: \ppquote{it was easier to type more into the prompt rather than to hold down the button itself}{P156}.

Some felt like they \ppquote{hadn't written enough prompt words}{P109} when they experienced a delay. For others, the time delay helped them to \ppquote{reflect on the prompt}{P116} and \ppquote{make adjustments}{P105} before it was submitted. This reflection may have encouraged creative thinking, supported by comments like: \ppquote{it got me to think more of how I wanted the story to be and pushed me to be more creative}{P40}.

\subsection{Summary}
Overall, our results suggest that a time delay can nudge people to write longer prompts. 
Although participants initially tried to submit shorter prompts in the first trial, they eventually submitted prompts that were significantly longer, and repeated this behaviour for subsequent trials. Some participants cannot or do not need to be nudged, but many who might normally write shorter prompts can be encouraged to write more. 

\section{Discussion}
Our first two experiments show that writing with a chat-based generative AI assistant lowers feelings of psychological ownership. These results extend and support prior work studying psychological ownership when writing with other human collaborators and early work examining writing with other forms of AI. The most profound finding from our work is a strong indication that when writing with a chat-based generative AI assistant, psychological ownership can be improved by requiring or encouraging users to write longer prompts. Specifically, writing longer prompts encouraged users to provide more details about the story plot, exerting more control over the output and making it more similar to their original prompt. This often required more mental demand and effort, which are important contributing factors to psychological ownership. To encourage users to write longer prompts for increased psychological ownership, we propose a simple change to the prompt submission button: the user must press and hold the button for a delay time if their prompt is too short. Our results show that this simple idea can increase prompt length. We believe that our work has the potential to change how we work and collaborate with generative AI assistants for increased psychological ownership.

\subsection{Applicability to Other Contexts}
Our experimental setup focused on a creative writing context, but our findings and interaction technique could be adapted to suit other writing contexts and even be beneficial for other outcomes beyond psychological ownership.

\subsubsection{Other Writing Contexts}
People may feel differently about owning different types of text. For example,
Nicholes \cite{nicholes2017measuring} found that undergraduate students feel more psychological ownership towards creative writing while graduate students feel more psychological ownership towards academic writing. Writers may also have different priorities and goals, with some valuing productivity and financial gain over creative fulfillment \cite{Biermann2022OwnershipLLM, li2024valueAIOwnership}. Similarly, prior work shows that writers benefit from receiving generated suggestions upon request to overcome writer's block \cite{TimedSuggestions2015} and from receiving feedback from collaborators earlier in the writing process \cite{Kumaran2021FeedbackEarly}. Generative AI writing assistants can be used as a tool to optimize for productivity and to receive suggestions and feedback, so depending on the goal, reduced psychological ownership may be a worthwhile trade-off for writers. Repeating a similar experiment but for a wider range of document types and writer goals would give insights into these trade-offs. In cases where psychological ownership is less important to writers, shorter prompts may suffice. One possibility is to further modify the interface to elicit goals from writers \cite{Kumaran2021FeedbackEarly} before using the generative AI assistant, and use these goals to determine whether the system should invoke techniques to encourage longer prompts from the writer. As the writer's goals change, such techniques could dynamically adapt to better suit them. For example, if the user's goal is to write an abstract, the system could require a longer prompt, but this requirement could be lifted or adjusted if the user's goal changes to receive suggestions for a specific sentence.

\subsubsection{Beyond Psychological Ownership}
Encouraging or requiring longer prompts could be beneficial to encourage other positive outcomes for users. For example, there are growing concerns that \emph{``effortlessly generated information''} provided by generative AI assistants will reduce critical thinking and problem-solving skills among students \cite{KASNECI2023102274}. Encouraging or enforcing longer prompts that increase the mental demand and effort could potentially mitigate such risks. Specifically, requiring students to write longer prompts could encourage them to put more thought and effort into core learning outcomes before receiving output from a generative AI assistant (e.g., thinking about the arguments and structure of an essay, or thinking about a coding algorithm).
Studying the effect of prompt length on other outcomes, like student learning outcomes, is an interesting avenue for future work.

\subsection{Implementation Approaches and Challenges}
We show that longer prompts can be encouraged through a simple user interface adjustment: adding a submission button time delay. However, there are related design considerations and possibilities for other approaches.

\subsubsection{Identifying Optimal ``Word Minimums''}
Our piloted interaction techniques all attempted to nudge users towards a known word minimum of 150 words as our first two experiments suggested that psychological ownership plateaued at 150 words for the given task. This 150-word minimum likely does not apply to other types of tasks and user goals, necessitating additional research and methods for identifying prompt lengths that optimize for psychological ownership.  
Another option is to create interaction techniques and interface augmentations that do not require any knowledge of optimal word minimums. For example, as the user writes, the generative AI assistant could dynamically append a sentence fragment to the prompt to elicit more information (e.g., ``Here is some more background information:'').

\subsubsection{Other Nudging Techniques}
We piloted several modifications to the user interface (Appendix \ref{section:appendixInteractionTechs}) before investigating the use of time delays to nudge users to write longer prompts. This approach requires minimal changes and can easily be integrated into current chat-based generative AI interfaces. However, other approaches may also encourage longer prompts and may be more effective than time delays. For example, the prompt entry interface could enforce a different prompt structure through scaffolding techniques \cite{Luther2015Scaffolding} (e.g., by breaking the prompt into multiple but smaller responses or requiring the user to ``fill in the blanks'' of a pre-defined prompt).

\subsubsection{Non-Chat-Based Interfaces}
Encouraging users to write longer prompts is immediately relevant to chat-based generative AI services like ChatGPT, but as AI becomes more integrated through techniques like direct manipulation \cite{masson2023directgpt}, factors other than prompt length may become more dominant. For example, sketches convey more information than text \cite{haught2003creativity}, and sketching can be an effective prompting technique \cite{yen2025codeshapingiterativecode}. Future work could explore the role of sketching to elicit more details from users for increased psychological ownership, which could benefit writing and even image generation tasks. 
An important implication of our work is that designers should think of ways to encourage users to invest more when interacting with generative AI assistants for increased psychological ownership.

\subsection{Limitations}

We chose to explore the role of prompt length within a single prompt, where the user mimics the role of a writing ``consultant'' \cite{Posner1992WriteTogetherStrategies}. In practice, users likely refine generated text over multiple prompts. A natural extension of our work is to explore the effect of word count across multiple messages within a conversation. This is representative of a broader class of user interactions with systems like ChatGPT, where users iterate over the output through multiple prompts.

One possible reason for the plateauing psychological ownership identified in Experiment 2 may be plateauing semantic text similarity. As models continue to improve to incorporate all details specified in prompts, this plateauing effect may disappear, necessitating further investigations of this effect for newer models. We did test if newer GPT-4 Turbo and 4-o models generated more similar output using the same prompts entered by our participants, but we did not observe any differences in \m{Text Similarity} between the older and newer models.

Another approach to improve psychological ownership is to design systems that incorporate all details indicated by users and that encourage humans to exert more control over AI-generated output (e.g., by manually accepting suggestions \cite{Lehmann2022SuggestionList}). We chose to implement and investigate systems that work like existing, widely-used commercial tools such as ChatGPT for increased ecological validity. However, as other generative AI systems that are designed to increase psychological ownership become more common, our results could serve as a valuable baseline for future studies, or could even be integrated into new systems for even higher feelings of psychological ownership.

The results from Experiment 3 may be lacking in ecological validity. Our experiment did not provide additional rewards to participants who wrote longer prompts, however, the effects of a time delay on prompt length may have been more pronounced due to the use of crowdworkers on Prolific, who are known to be highly conscientious participants \cite{douglas2023data}. As such, they may have been more willing to write longer prompts to do the task ``well'' and less likely to abandon using such systems than the general public. Similarly, the 20- and 60-second time delays worked well in the context of this experiment, but they may not transfer well to real-world systems. It is possible that even shorter time delays, such as 2-second delays studied in prior work \cite{nah2004study, SchneidermanResponseTimes, Lewis2020Delays}, may also increase prompt length while mitigating risks of users abandoning the system \cite{KopsellDropoffDelay}. Providing users with feedback and detailed explanations on why a delay exists can also increase their willingness to wait \cite{nah2004study, egelman2010please}. Deploying variations of this nudging technique in real-world systems would give more insights into optimal design parameters that strike a balance between increasing prompt length and preventing system abandonment.

\section{Conclusion}
Our work shows that writing with an AI assistant reduces feelings of psychological ownership, but requiring or encouraging users to write longer prompts can improve it. This is no replacement for writing something independently, but it is a simple way to encourage writers to include more details about the content and put more thought and effort into their prompts. A time delay experienced when submitting a prompt can effectively increase prompt length along with psychological ownership, a simple interface change to positively impact human-AI collaborations. 

\begin{acks}
This work was made possible by 
NSERC Discovery Grant 2024-03827.
\end{acks}

\bibliographystyle{ACM-Reference-Format}
\bibliography{main_acm.bib}

\appendix
\newpage
\onecolumn
\section{Experiment 1: Analysis Details}
\label{section:appendixStudy1}
\renewcommand\thetable{\thesection.\arabic{table}}
\setcounter{table}{0}
\renewcommand\thefigure{\thesection.\arabic{figure}}
\setcounter{figure}{0}

\begin{table*}[h!]
    \centering
    \caption{Demographic information for Experiment 1, adapted from Masson et al. \cite{masson2023directgpt}.}
    \small %
\begin{tabular}{lr|lr|lr|lr|lr}
\toprule
\multicolumn{2}{l|}{Gender} & \multicolumn{2}{l|}{Age} & \multicolumn{2}{l|}{Education} & \multicolumn{2}{l|}{Creative Writing Frequency} & \multicolumn{2}{l}{Prompt Engineering Familiarity}\\
\midrule
Men & 14 & 25-34 & 10 & Some University (no credit) & 4 & Daily & 1 & Extremely & 3\\
Women & 16 & 35-44 & 7 & Technical Degree & 1 & Weekly & 7 & Moderately & 3\\
Non-binary & 1 & 45-54 & 6 & Bachelor's Degree & 21 & Monthly & 7 & Somewhat & 3 \\
&& 55-64 & 6 & Master's Degree & 5 & Less than Monthly & 11 & Slightly & 7\\
&&65-74& 2&  &  & Never & 5 & Not at All & 15\\
\end{tabular}

\begin{tabular}{lr|lr|lrlr|lrlr}
\\
\toprule
\multicolumn{2}{l|}{ChatGPT Frequency} & \multicolumn{2}{l|}{Weekly ChatGPT Use} & \multicolumn{4}{l|}{Other AI Services} & \multicolumn{4}{l}{AI Usage}\\
\midrule
Daily & 3 & 0 times & 7 & Nothing & 3 & Gemini & 9 & Nothing & 1 & Generate Images & 12\\
Weekly & 14 & 1-5 times & 13 & GPT-2 & 6 & Grammarly & 9 & Write Emails & 11 & Get a Definition & 9\\
Monthly & 4 & 5-10 times & 8 & GPT-3 & 12 & Stable Diffusion & 2 & Write Papers/Essays & 2 & Explore a Topic & 17\\
Less than Monthly & 9 & 10-30 times & 2 & GPT-4 & 6 & Dall-E & 4 & Write Stories & 6 & Brainstorming & 25\\
Never & 1 & 30+ times & 1 & Copilot & 2 & Midjourney & 5 & Edit Text & 14 & Find References & 8 \\
&&&& Bing & 12 & Other & 6 & Generate Code & 3 & Clarification & 7\\
&&&&&&&& Edit Code & 3 & Translate Text & 5\\
&&&&&&&& Debug Code & 3 & Other & 4\\
\end{tabular}
    \Description{Demographic information: gender, age, education, creative writing frequency, prompt engineering familiarity, ChatGPT frequency, weekly ChatGPT use, other AI services, and AI usage. The demographics cover a diverse range.}
    \label{tab:exp1Demo}
\end{table*}

\begin{table*}[h!]
    \centering
    \caption{Omnibus and post hoc statistical tests for Experiment 1: (a) \m{Psychological Ownership}, (b) \m{Duration}, (c) \m{Word Count}, (d) \m{Text Similarity}, (e) \m{Mental Demand},  and (f) \m{Effort}.}
    
\newcommand{\tabspacelg}{\vspace{0.8em}}
\newcommand{\tabspacesm}{\vspace{0.25em}}

\small
\begin{tabular}{llrr|llrr|llrr}

\toprule
\multicolumn{4}{l}{\tabspacesm (a) \m{Psychological Ownership}} & \multicolumn{4}{l}{\tabspacesm (b) \m{Duration}} & \multicolumn{4}{l}{\tabspacesm (c) \m{Word Count}}\\
\multicolumn{4}{l}{\tabspacesm \friedmanN{3}{31}{78.01}{.001}{.84}} &  \multicolumn{4}{l}{\tabspacesm \friedmanN{3}{31}{35.86}{.001}{.38}} & \multicolumn{4}{l}{\tabspacesm \friedmanN{3}{31}{85.87}{.001}{.92}}\\
\multicolumn{2}{l}{\textit{comparisons}} & \multicolumn{2}{l}{\textit{p-value}} & \multicolumn{2}{l}{\textit{comparisons}} & \multicolumn{2}{l}{\textit{p-value}} & \multicolumn{2}{l}{\textit{comparisons}} & \multicolumn{2}{l}{\textit{p-value}}  \\ 
\midrule
\f{ai-3} & \f{ai-50} & < .001 & *** & \f{ai-3} & \f{ai-50} & .87 & \textit{n.s.} & \f{ai-3} & \f{ai-50} & < .001 & *** \\ 
\f{ai-3} & \f{ai-150} & < .001 & *** & \f{ai-3} & \f{ai-150} & .02 & * & \f{ai-3} & \f{ai-150} & < .001 & *** \\ 
\f{ai-3} & \f{no-ai} & < .001 & *** & \f{ai-3} & \f{no-ai} & .01 & ** & \f{ai-3} & \f{no-ai} & < .001 & *** \\ 
\f{ai-50} & \f{ai-150} & .01 & ** & \f{ai-50} & \f{ai-150} & < .001 & ***  & \f{ai-50} & \f{ai-150} & < .001 & ***\\
\f{ai-50} & \f{no-ai} & < .001 & *** & \f{ai-50} & \f{no-ai}  & < .001 & *** & \f{ai-50} & \f{no-ai} & < .001 & *** \\
\f{ai-150} & \f{no-ai} & < .001 & *** & \f{ai-150} & \f{no-ai} & .24 & \textit{n.s.} & \f{ai-150} & \f{no-ai} & .005 & **\\
\end{tabular}

\begin{tabular}{llrr|llrr|llrr}
\toprule
\multicolumn{4}{l}{\tabspacesm (d) \m{Text Similarity}} & \multicolumn{4}{l}{\tabspacesm (e) \m{Mental Demand}} & \multicolumn{4}{l}{\tabspacesm (f) \m{Effort}}\\
\multicolumn{4}{l}{\tabspacesm \friedmanN{3}{31}{99.00}{.001}{1.00}} &  \multicolumn{4}{l}{\tabspacesm \friedmanN{3}{31}{60.21}{.001}{.65}} & \multicolumn{4}{l}{\tabspacesm \friedmanN{3}{31}{49.99}{.001}{.54}} \\
\multicolumn{2}{l}{\textit{comparisons}} & \multicolumn{2}{l}{\textit{p-value}} & \multicolumn{2}{l}{\textit{comparisons}} & \multicolumn{2}{l}{\textit{p-value}} & \multicolumn{2}{l}{\textit{comparisons}} & \multicolumn{2}{l}{\textit{p-value}} \\ 
\midrule
\f{ai-3} & \f{ai-50} & < .001 & *** & \f{ai-3} & \f{ai-50} & < .001 & *** & \f{ai-3} & \f{ai-50} & < .001 & *** \\ 
\f{ai-3} & \f{ai-150} & < .001 & *** & \f{ai-3} & \f{ai-150} & < .001 & *** & \f{ai-3} & \f{ai-150} & < .001 & *** \\ 
\f{ai-3} & \f{no-ai} & < .001 & *** & \f{ai-3} & \f{no-ai} & < .001 & *** & \f{ai-3} & \f{no-ai} & < .001 & *** \\ 
\f{ai-50} & \f{ai-150} & < .001 & *** & \f{ai-50} & \f{ai-150} & < .001 & **  & \f{ai-50} & \f{ai-150} & .16 & \textit{n.s.}\\
\f{ai-50} & \f{no-ai} & < .001 & *** & \f{ai-50} & \f{no-ai}  & < .001 & *** & \f{ai-50} & \f{no-ai} & .02 & * \\
\f{ai-150} & \f{no-ai} & < .001 & *** & \f{ai-150} & \f{no-ai} & .33 & \textit{n.s.} & \f{ai-150} & \f{no-ai} & .13 & \textit{n.s.} \\
\bottomrule

\end{tabular}

    \Description{Details of statistical tests for Psychological Ownership, Duration, Word Count, Text Similarity, Mental Demand, and Effort. All omnibus tests are significant and most post hoc tests are significant.}
    \label{tab:exp1Significance}
\end{table*}

\clearpage
\newpage
\section{Experiment 2: Analysis Details}
\setcounter{table}{0}
\setcounter{figure}{0}
\label{section:appendixStudy2}

\begin{table*}[h!]
    \centering
    \caption{Demographic information for Experiment 2.}
    \Description{Demographic information: gender, age, education, creative writing frequency, prompt engineering familiarity, ChatGPT frequency, weekly ChatGPT use, other AI services, and AI usage. The demographics cover a diverse range.}
    \small %
\begin{tabular}{lr|lr|lr|lr|lr}
\toprule
\multicolumn{2}{l|}{Gender} & \multicolumn{2}{l|}{Age} & \multicolumn{2}{l|}{Education} & \multicolumn{2}{l|}{Creative Writing Frequency} & \multicolumn{2}{l}{Prompt Engineering Familiarity}\\
\midrule
Men & 12 & 18-24 & 3 & High School Diploma & 3 & Daily & 2 & Extremely & 1\\
Women & 19 & 25-34 & 11 & Some University (no credit) & 8 & Weekly & 3 & Moderately & 3\\
Non-binary & 1 & 35-44 & 4 & Technical Degree & 2 & Monthly & 7 & Somewhat & 7 \\
Other & 1 & 45-54 & 11 & Bachelor's Degree & 18 & Less than Monthly & 12 & Slightly & 12\\
Unknown & 1 & 55-64 & 4 & Master's Degree & 2 & Never & 10 & Not at All & 11\\
& & Unknown & 1 & Doctorate Degree & 1
\end{tabular}

\begin{tabular}{lr|lr|lrlr|lrlr}
\\
\toprule
\multicolumn{2}{l|}{ChatGPT Frequency} & \multicolumn{2}{l|}{Weekly ChatGPT Use} & \multicolumn{4}{l|}{Other AI Services} & \multicolumn{4}{l}{AI Usage}\\
\midrule
Daily & 10 & 0 times & 8 & Nothing & 6 & Gemini & 7 & Nothing & 2 & Generate Images & 14\\
Weekly & 8 & 1-5 times & 12 & GPT-2 & 6 & Grammarly & 8 & Write Emails & 14 & Get a Definition & 17\\
Monthly & 6 & 5-10 times & 4 & GPT-3 & 12 & Stable Diffusion & 3 & Write Papers/Essays & 4 & Explore a Topic & 22\\
Less than Monthly & 6 & 10-30 times & 6 & GPT-4 & 6 & Dall-E & 9 & Write Stories & 5 & Brainstorming & 19\\
Never & 4 & 30+ times & 4 & Copilot & 0 & Midjourney & 6 & Edit Text & 15 & Find References & 5 \\
&&&& Bing & 10 & Other & 4 & Generate Code & 3 & Clarification & 16\\
&&&&&&&& Edit Code & 3 & Translate Text & 5\\
&&&&&&&& Debug Code & 3 & Other & 3\\
\end{tabular}
    \label{tab:exp2Demo}
\end{table*}

\begin{table*}[h!]
    \centering
    \caption{Omnibus and post hoc statistical tests for Experiment 2: (a) \m{Psychological Ownership}, (b) \m{Duration}, (c) \m{Word Count}, and (d) \m{Text Similarity}. Note that \m{Mental Demand} and \m{Effort} are excluded as these omnibus tests were not significant.}
    \Description{Details of statistical tests for Psychological Ownership, Duration, Word Count, and Text Similarity. All omnibus tests are significant. All but one post hoc tests are significant for Word Count. For all other metrics, the only significant differences are between the AI conditions and NO-AI.}
    
\newcommand{\tabspacelg}{\vspace{0.8em}}
\newcommand{\tabspacesm}{\vspace{0.25em}}

\small
\begin{tabular}{llrr|llrr|llrr|llrr}

\toprule
\multicolumn{4}{l}{\tabspacesm (a) \m{Psychological Ownership}} & \multicolumn{4}{l}{\tabspacesm (b) \m{Duration}} & \multicolumn{4}{l}{\tabspacesm (c) \m{Word Count}} & \multicolumn{4}{l}{\tabspacesm (d) \m{Text Similarity}}\\
\multicolumn{4}{l}{\tabspacesm \friedmanN{3}{34}{43.63}{.001}{.43}} &  \multicolumn{4}{l}{\tabspacesm \friedmanN{3}{34}{11.32}{.01}{.11}} & \multicolumn{4}{l}{\tabspacesm \friedmanN{3}{34}{71.27}{.001}{.70}} & \multicolumn{4}{l}{\tabspacesm \friedmanN{3}{34}{61.94}{.001}{.61}}\\
\multicolumn{2}{l}{\textit{comparisons}} & \multicolumn{2}{l}{\textit{p-value}} & \multicolumn{2}{l}{\textit{comparisons}} & \multicolumn{2}{l}{\textit{p-value}} & \multicolumn{2}{l}{\textit{comparisons}} & \multicolumn{2}{l}{\textit{p-value}} & \multicolumn{2}{l}{\textit{comparisons}} & \multicolumn{2}{l}{\textit{p-value}} \\ 
\midrule
\f{ai-150} & \f{ai-175} & .69 & \textit{n.s.} & \f{ai-150} & \f{ai-175} & 1.0 & \textit{n.s.} & \f{ai-150} & \f{ai-175} & < .001 & *** & \f{ai-150} & \f{ai-175} & .46 & \textit{n.s.}\\ 
\f{ai-150} & \f{ai-200} & .69 & \textit{n.s.} & \f{ai-150} & \f{ai-200} & .03 & * & \f{ai-150} & \f{ai-200} & < .001 & *** & \f{ai-150} & \f{ai-200} & .46 & \textit{n.s.} \\ 
\f{ai-150} & \f{no-ai} & < .001 & *** & \f{ai-150} & \f{no-ai} & 1.0 & \textit{n.s.} & \f{ai-150} & \f{no-ai} & .16 & \textit{n.s.} & \f{ai-150} & \f{no-ai} & < .001 & ***\\ 
\f{ai-175} & \f{ai-200} & .69 & \textit{n.s.} & \f{ai-175} & \f{ai-200} & .14 & \textit{n.s.}  & \f{ai-175} & \f{ai-200} & < .001 & *** & \f{ai-175} & \f{ai-200} & .77 & \textit{n.s.}\\
\f{ai-175} & \f{no-ai} & < .001 & *** & \f{ai-175} & \f{no-ai}  & 1.0 & \textit{n.s.} & \f{ai-175} & \f{no-ai} & .003 & ** & \f{ai-175} & \f{no-ai} & < .001 & *** \\
\f{ai-200} & \f{no-ai} & < .001 & *** & \f{ai-200} & \f{no-ai} & .13 & \textit{n.s.} & \f{ai-200} & \f{no-ai} & < .001 & *** & \f{ai-200} & \f{no-ai} & < .001 & ***\\
\bottomrule
\end{tabular}

    \label{tab:exp2Significance}
\end{table*}

\clearpage
\newpage
\section{Preliminary Interface Modifications and Interaction Techniques}
\setcounter{table}{0}
\setcounter{figure}{0}
\label{section:appendixInteractionTechs}

\begin{figure*}[t]
	\centering
	\includegraphics[width=\textwidth]{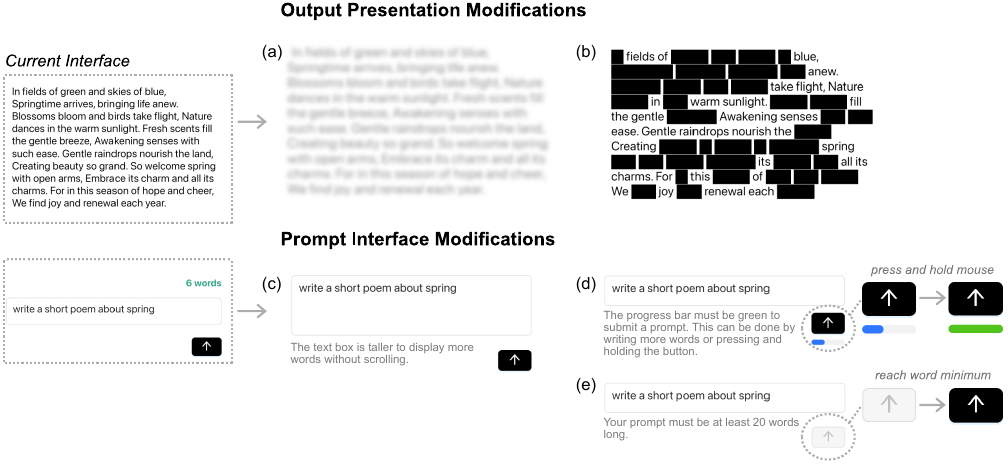}
	\caption{Prompt entry interface designs that augment the prompt entry area and the response area: (a) blurring or (b) redacting the text more if the prompt contained less words; or nudging users to write longer prompts by (c) making the prompt entry text box taller or (d) requiring the user to press and hold the submit button. We compared these to the baseline ``hard'' constraint of (e) preventing prompt submission if the prompt is not long enough. 
    }
    \Description{Current chat-based AI interface on the left, with the output on the top and the text entry on the bottom. To the right are modifications to the output and the text entry input. Output modifications include blurred text and text that has some words redacted with black rectangles. Input modifications include increasing the height of the text box, disabling a Submit button until the word minimum has been reached, or making the user hold their cursor over a button to move a progress bar. All input modifications have text underneath classifying the type of document being requested ("assuming a poem between 20 and 100 words, your prompt should be at least 15 words").}
	\label{fig:interactions}
\end{figure*}

Before deciding to explore the use of time delays to increase prompt length in more depth, we also considered the following modifications to both the output presentation and the prompt entry interfaces (Figure \ref{fig:interactions}). These modifications were the result of brainstorming sessions between the authors, and were selected as they are compatible with current chat-based generative AI tools like ChatGPT. %
\begin{itemize}
    \item \textbf{Increasingly blurring the generated output} the shorter their prompt is, requiring the user to press and hold it make the output clearer and easier to read (Figure \ref{fig:interactions}a). Much like the time delay technique, the time spent pressing and holding the output is mapped to prompt length. This technique was inspired by Cockburn et al.'s frost-brushing technique for teaching users about the spatial information of user interfaces \cite{cockburn2007FrostBrush}.
    \item \textbf{Increasingly redacting parts of the generated output} the shorter their prompt is, ``punishing'' the user if the prompt length was too short (Figure \ref{fig:interactions}b). If the user writes a shorter prompt, each chunk of generated output has a higher chance of being redacted. Having a few chunks redacted unlikely hinders overall comprehension as the user can easily ``fill in the blanks,'' but if a majority of the text is redacted and the output is illegible, the user has to increase the length of their existing prompt to see more of the generated text next time.
    \item \textbf{Increasing the height of the prompt entry text box} so it can fully hold the word minimum (150 words).
\end{itemize}

\begin{figure*}[h!]
\centering
	\includegraphics[width=0.47\textwidth]{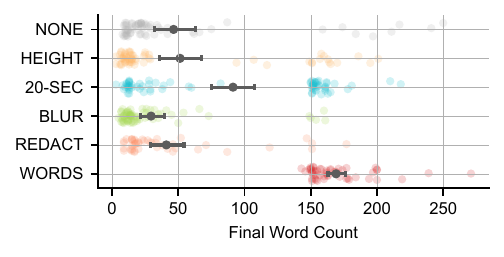}
	\caption{\m{Final Word Count} by \f{condition} for the initial pilot experiment with all interface modifications.}
    \Description{Confidence intervals and individual participant data for Final Word Count for the initial pilot experiment. There are two clusters of data points for 20-SEC: one between 0 and 50 words, and another at 150 words.}
	\label{fig:pilotExp3} %
\end{figure*}

\newpage
\section{Experiment 3: Analysis Details}
\setcounter{table}{0}
\setcounter{figure}{0}
\label{section:appendixStudy3}

\begin{table*}[h!]
    \centering
    \caption{Demographic information for Experiment 3.}
    \small %
\begin{tabular}{lr|lr|lr|lr|lr}
\toprule
\multicolumn{2}{l|}{Gender} & \multicolumn{2}{l|}{Age} & \multicolumn{2}{l|}{Education} & \multicolumn{2}{l|}{Creative Writing Frequency} & \multicolumn{2}{l}{Prompt Engineering Familiarity}\\
\midrule
Men & 93 & 18-24 & 5 & Less than a High School Diploma & 2 & Daily & 2 & Extremely & 8\\
Women & 61 & 25-34 & 47 & High School Diploma & 22 & Weekly & 16 & Moderately & 26\\
Non-binary & 2 & 35-44 & 51 & Some University (no credit) & 37 & Monthly & 31 & Somewhat & 28 \\
&& 45-54 & 25 & Technical Degree & 9 & Less than Monthly & 73 & Slightly & 44\\
&&55-64& 16 & Bachelor's Degree & 65 & Never & 34 & Not at All & 50\\
&&65-74 & 11 & Professional Degree Beyond Bachelor's Degree & 1\\
&&75+ & 1 & Master's Degree & 20\\
\end{tabular}

\begin{tabular}{lr|lr|lrlr|lrlr}
\\
\toprule
\multicolumn{2}{l|}{ChatGPT Frequency} & \multicolumn{2}{l|}{Weekly ChatGPT Use} & \multicolumn{4}{l|}{Other AI Services} & \multicolumn{4}{l}{AI Usage}\\
\midrule
Daily & 19 & 0 times & 34 & Nothing & 12 & Gemini & 65 & Nothing & 4 & Generate Images & 68\\
Weekly & 76 & 1-5 times & 70 & GPT-2 & 37 & Grammarly & 54 & Write Emails & 52 & Get a Definition & 59\\
Monthly & 27 & 5-10 times & 32 & GPT-3 & 79 & Stable Diffusion & 12 & Write Papers/Essays & 21 & Explore a Topic & 111\\
Less than Monthly & 26 & 10-30 times & 16 & GPT-4 & 60 & Dall-E & 36 & Write Stories & 35 & Brainstorming & 100\\
Never & 8 & 30+ times & 4 & Copilot & 14 & Midjourney & 19 & Edit Text & 65 & Find References & 35 \\
&&&& Bing & 82 & Other & 13 & Generate Code & 26 & Clarification & 67\\
&&&&&&&& Edit Code & 19 & Translate Text & 33\\
&&&&&&&& Debug Code & 15 & Other & 11\\
\end{tabular}
    \Description{Demographic information: gender, age, education, creative writing frequency, prompt engineering familiarity, ChatGPT frequency, weekly ChatGPT use, other AI services, and AI usage. The demographics cover a diverse range.}
    \label{tab:exp3Demo}
\end{table*}

\begin{table*}[h!]
    \centering
    \caption{Omnibus and post hoc statistical tests for Experiment 3: (a) \m{Final Word Count}, (b) \m{Submission Attempts}, (c) \m{Attempted Word Count} for \f{trial} = \f{1}.}
    \Description{Details of statistical tests for Final Word Count, Submission Attempts, and Attempted Word Count. All omnibus tests are significant.}
    \newcommand{\tabspacelg}{\vspace{0.8em}}
\newcommand{\tabspacesm}{\vspace{0.25em}}

\newcommand{\art}[5]{{\small$F_{#1,#2}=#3$, $p<#4$, $\eta_P^2=#5$}}
\small
\begin{tabular}{llrr|llrr}

\toprule
\multicolumn{8}{c}{(a) \m{Final Word Count}}\\
\multicolumn{4}{l}{\tabspacesm \f{condition} (\art{4}{151}{26.45}{.001}{0.41})} & \multicolumn{4}{l}{\tabspacesm \f{trial} (\art{3}{453}{2.87}{.05}{0.02})} \\  
\multicolumn{2}{l}{\textit{comparisons}} & \multicolumn{2}{l}{\textit{p-value}}\\ 
\midrule
\f{none} & \f{0-sec} & 1.00 & \textit{n.s.} & \f{1} & \f{2} & .11 & \textit{n.s.}\\ 
\f{none} & \f{20-sec} & .042 & * & \f{1} & \f{3} & .13 & \textit{n.s.}\\
\f{none} & \f{60-sec} & .013 & * & \f{1} & \f{4} & .049 & *\\
\f{none} & \f{words} & < .001 & *** & \f{2} & \f{3} & 1.00 & \textit{n.s.}\\
\f{0-sec} & \f{20-sec} & .013 & * & \f{2} & \f{4} & 1.00 & \textit{n.s.}\\
\f{0-sec} & \f{60-sec} & .0024 & ** & \f{3} & \f{4} & 1.00 & \textit{n.s.}\\
\f{0-sec} & \f{words} & < .001 & ***\\
\f{20-sec} & \f{60-sec} & 1.00 & \textit{n.s.}\\
\f{20-sec} & \f{words} & < .001 & ***\\
\f{60-sec} & \f{words} & < .001 & ***\tabspacelg\\

\toprule
\multicolumn{4}{c}{(b) \m{Submission Attempts}} & \multicolumn{4}{c}{(c) \m{Attempted Word Count}}\\
\multicolumn{4}{l}{\tabspacesm \f{trial} (\art{3}{183}{96.95}{.001}{0.61})} & \multicolumn{4}{l}{\tabspacesm \f{condition} (\friedmanEta{4}{156}{71.11}{.001}{.17})} \\ 
\multicolumn{2}{l}{\textit{comparisons}} & \multicolumn{2}{l}{\textit{p-value}}  & \multicolumn{2}{l}{\textit{comparisons}} & \multicolumn{2}{l}{\textit{p-value}}   \\  
\midrule
\f{1} & \f{2} & < .001 & *** & \f{none} & \f{0-sec} & 1.00 & \textit{n.s.}\\ 
\f{1} & \f{3} & < .001 & *** & \f{none} & \f{20-sec} & 1.00 & \textit{n.s.}\\
\f{1} & \f{4} & < .001 & *** & \f{none} & \f{60-sec} & 1.00 & \textit{n.s.}\\
\f{2} & \f{3} & .15 & \textit{n.s.} & \f{none} & \f{words} & < .001 & ***\\
\f{2} & \f{4} & .84 & \textit{n.s.} & \f{0-sec} & \f{20-sec} & 1.00 & \textit{n.s.}\\
\f{3} & \f{4} & .16 & \textit{n.s.} & \f{0-sec} & \f{60-sec} & 1.00 & \textit{n.s.}\\
&&&& \f{0-sec} & \f{words} & < .001 & ***\\
&&&& \f{20-sec} & \f{60-sec} & 1.00 & \textit{n.s.}\\
&&&& \f{20-sec} & \f{words} & < .001 & ***\\
&&&& \f{60-sec} & \f{words} & < .001 & ***\tabspacelg\\

\bottomrule

\end{tabular}

    \label{tab:exp3Significance}
\end{table*}

\end{document}